\documentclass[showpacs,preprintnumbers]{revtex4}
\usepackage{amssymb}
\usepackage{amsmath}
\usepackage{graphicx}
\usepackage{dcolumn}
\usepackage{bm}
\usepackage{epsf}

\begin{document}
\title{Some applications of the Faddeev-Yakubovsky equations to the cold-atom physics}

\author{Jaume Carbonell}
\email{carbonell@lpsc.in2p3.fr}
\affiliation{Laboratoire Physique Subatomique et Cosmologie, Universit\'e Jospeh Fourier, CNRS/IN2P3\\
 53, avenue des Martyrs, 38026 Grenoble Cedex, France.}
\author{Arnoldas Deltuva}
\email{deltuva@cii.fc.ul.pt}
\affiliation{Centro de F\'{\i}sica Nuclear da Universidade de Lisboa, P-1649-003 Lisboa, Portugal}
\author{Rimantas Lazauskas}
\email{rimantas.lazauskas@ires.in2p3.fr}
\affiliation{IPHC, IN2P3-CNRS/Universit\'e Louis Pasteur BP 28,
F-67037 Strasbourg Cedex 2, France}

\date{\today }
\pacs{PACS number}

\begin{abstract}
We  present some recent applications of the Faddeev--Yakubovsky  equations
in describing atomic bound and scattering problems. We consider
the scattering of a charged particle $X$  by atomic hydrogen with special interest in $X=p,e^{\pm}$,
systems of cold bosonic molecules and the bound and scattering properties  of N=3 and N=4 atomic $^4$He multimers.
\end{abstract}

\maketitle

\section{Introduction}\label{sec:Intro}

The Faddeev-Yakubovsky (FY) equations  constitute a rigorous mathematical formulation
of the quantum mechanical N-body problem  in the framework of non relativistic dynamics.
They allow obtaining exact solutions - in the numerical sense -
of the Schr\"{o}dinger equation for bound and scattering states, in principle for an arbitrary number of particles.
They were first formulated  as integral equations in momentum space by Faddeev in the early sixties~\cite{Fad_60},
 in the context of 3-nucleon problem
with short range interactions, and were generalized some years latter to an arbitrary number  (N) of particles by Yakubovsky~\cite{Yaku_67}.

A substantial progress in their numerical solution was made when the boundary conditions in the configuration space  were established for N=3 in~\cite{Merkuriev_71,Merkuriev_74,MGL_76}  and for N$>$3 in~\cite{MY_83}.
A reformulation of the original Faddeev equations, allowing to incorporate long range Coulomb like interactions was derived in~\cite{Merkuriev_80,Merkuriev_81}.
A brief review of the different steps in elaborating the Faddeev-Yakubovsky  approach to N-body problem can be found in~\cite{Moto_2008}.

Until now only  N=3 and N=4 systems have been explored, although there is no any formal nor practical impediment to go beyond.
Nowadays very accurate FY solutions have been obtained for the bound state problems on hadronic, nuclear and atomic systems, in configuration as well as in momentum
space~\cite{MYG_84,NKG_PRL85_00,LC_PRA73_2006,LC_PRC70_04,Papp_PRL94_05}, still with a smaller accuracy   than the variational methods.
The real advantage in using the FY equations, which was also their original motivation, is in describing on the same foot the  scattering states.
They play there an unavoidable role, especially when many channels are open including the break-up ones  (dissociation into more than two clusters).
At present, there exist three-body reliable scattering results in nuclear and atomic physics including the break-up
process~\cite{GWH_PR_96,PHH_PRA63_01,KCG_PRA_92,KCG_PRA_95, LC_FBS31_2002}.
In the four-body case, the progress remains limited to elastic and  rearrangement 1+3 and 2+2 particle channel in nuclear and atomic problems~\cite{FY_PAN63_00,CC_PRC58_98,LC_PRC70_04, LCFVKR_PRC71_05,LC_PRA73_2006,DF_PRC81_10}.

We present in this contribution  some recent applications  of the Faddeev-Yakubovsky formalism to describe cold atomic structures.
A short introduction to the Faddeev equations is given in Section~\ref{Formalism}.
Section~\ref{p_H} is devoted to present some results of  a charged particle interacting with atomic hydrogen, with special interest in the $p-H$ (or $H_2^+$ molecular ion)  and $e^{\pm}-H$ systems.
We study in Section~\ref{Cold}  universal properties of the weakly bound N=3 and N=4 cold bosonic molecules.
Some results concerning $^4$He atomic multimers are given in Section~\ref{He_mers}.
The last section is devoted to draw some conclusion and perspectives.

\section{The formalism}\label{Formalism}

We will develop in what follows the main ideas of the Faddeev equations for solving the three-body problem in configuration space.
This form is the best adapted to deal with atomic problems.
Let us consider three different  spinless particles with masses $m_i$, coordinates $\vec{r}_i$ and denote  by $(\vec{x}_{i},\vec{y}_i)$
the three different sets of Jacobi coordinates defined by
\begin{eqnarray}
\vec{x}_{i}&=&  \sqrt{2m_{j}m_{k}\over {m_0(m_{j}+m_{k})}}(\vec{r}_{j}-\vec{r}_k)  \hspace{2.5cm}i=1,2,3  \label{x_i}\\
\vec{y}_{i}&=&\sqrt{2m_{i}(m_{j}+m_{k})\over m_0M}\left[\vec{r}_{i}- {m_{j}\vec{r}_{j}+m_{k}\vec{r}_{k}\over m_{j}+m_{k}} \right]
\end{eqnarray}
where ($ijk$) denotes a cyclic permutation of the particle numbers (123), $M=m_1+m_2+m_3$ is the total mass of the system
and $m_0$ an arbitrary mass to be fixed at convenience.
We will also assume that these particles interact via pairwise short range local potentials $V_i(x_i)$ vanishing at some distance $R_i$.

\begin{figure}[h]
\begin{minipage}{7.5cm}
\begin{center}
\mbox{\epsfxsize=7.cm\epsffile{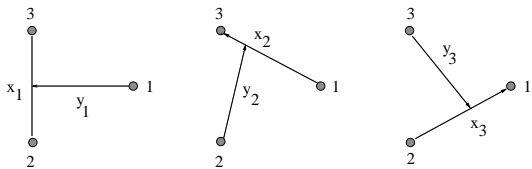}}
\caption{Sets of three-particle Jacobi coordinates.}\label{NR_R_LFD_BS}
\end{center}
\end{minipage}
\end{figure}

The dynamics of the system  can be alternatively described by  $\{\vec{r}_i\}_{i=1,2,3}$ or by choosing one of the Jacobi
coordinate sets $(\vec{x}_{i},\vec{y}_i)$, completed by the center of mass coordinate $\vec{R}$
\[ M\vec{R}= {m_1\vec{r}_1+m_2\vec{r}_2+m_3\vec{r}_3}\]
The three-body Hamiltonian is assumed to have the form
\begin{equation}\label{H}
Ê{\cal H}= {\cal H}_0 +V
\end{equation}
where ${\cal H}_0$ is the kinetic energy which can be written in one of these forms
\[{\cal H}_0 = - {\hbar^2\over2}  \left({1\over m_1}\Delta_{\vec{r}_1}+{1\over m_2}\Delta_{\vec{r}_2} +{1\over m_3}\Delta_{\vec{r}_3} \right)
           = - \frac{\hbar^2}{m_0}\left[\Delta_{\vec{x_i}}+\Delta_{\vec{y_i}}+{m_0\over2M}\Delta_{\vec{R}}\right] \]
and $V$ the total potential
\begin{equation}\label{V}
   V = V_1(x_1) + V_2(x_2)+V(x_3)
\end{equation}

In order to remove the center of mass motion, the choice of a Jacobi set is mandatory.  Only in terms of it,
the three-body wavefunction $\Psi$, eigenstate of the total three-body Hamiltonian~(\ref{H}), factorizes into
\[ \Psi(\vec{x}_i,\vec{y}_i,\vec{R} )= \Phi(\vec{x}_i,\vec{y}_i) \; e^{i\vec{P}\cdot\vec{R}}\]
where $\Phi(\vec{x}_i,\vec{y}_i)$ is an eigenstate of the intrinsic Hamiltonian
\begin{equation}\label{HI}
 H\Phi= E\Phi   \qquad   H= - \frac{\hbar^2}{m_0}\left[\Delta_{\vec{x_i}}+\Delta_{\vec{y_i}}  \right] +V
\end{equation}
the only one that we are going to consider from now.

As it can be already seen from $(\ref{V})$,  none of the Jacobi sets is neither privileged nor fully satisfactory.
The "non interacting" region of particles 2 and 3 is simply given by $x_1>R_1$ but it is difficult to
define this region in terms of  say ($\vec{x}_2,\vec{y}_2)$.

On the other hand, if we wish to describe for instance the scattering of particle 1 on a bound state of  particles 2 and 3 - denoted symbolically by 1(2,3) --
it will be natural to chose the coordinate set $(\vec{x}_1,\vec{y}_1)$.
However the final state can contain, together with the initial state, a superposition of  channels 2(1,3) and/or 3(1,2) which can be hardly described
in terms of the same coordinate set.
For the bound states and break-up channels the three ensembles will appear naturally on the same footing.

The Faddeev equations
are based on a partition of the total wave function on as many components as two-body asymptotic channels:
\[  \Phi = \Phi_1  + \Phi_2  + \Phi_3 \]
It is straightforward to see that the three-body Schr\"{o}dinger equation (\ref{HI}) is equivalent to the set
of coupled partial differential equations for the Faddeev components $\Phi_i$
\begin{eqnarray} \label{FE}
\left[ E-H_0-V_1(x_1) \right] \Phi_1(\vec{x}_1,\vec{y}_1) &=& V_1(x_1) \left[  \Phi_2(\vec{x}_2,\vec{y}_2)+ \Phi_3(\vec{x}_3,\vec{y}_3) \right]  \cr
\left[ E-H_0-V_2(x_2) \right] \Phi_2(\vec{x}_2,\vec{y}_2) &=& V_2(x_2) \left[  \Phi_3(\vec{x}_3,\vec{y}_3)+ \Phi_1(\vec{x}_1,\vec{y}_1) \right]  \cr
\left[ E-H_0-V_3(x_3) \right] \Phi_3(\vec{x}_3,\vec{y}_3) &=& V_3(x_3) \left[  \Phi_1(\vec{x}_1,\vec{y}_1)+ \Phi_2(\vec{x}_2,\vec{y}_2) \right]
\end{eqnarray}
The coupling is ensured by the right hand side. It is highly non local due to the linear relations
between two different sets of Jacobi coordinates: $\vec{x}_{\alpha}(\vec{x}_{\beta},\vec{y}_{\beta})$, $\vec{y}_{\alpha}(\vec{x}_{\beta},\vec{y}_{\beta})$.
In the non interacting region, $V_i=0$,  the three Faddeev equations decouple  and the boundary conditions
for each component take a simple form when expressed in their own Jacobi coordinate set.

These boundary conditions are more easily implemented in terms of the reduced Faddeev components $\phi_i$ defined by:
\begin{eqnarray}
\phi_i=x_i y_i \Phi_i
\end{eqnarray}
and take the following  Dirichlet form:
\begin{itemize}
\item[--] they vanish for $x_i=0$ and $y_i=0$:
\begin{eqnarray}
\phi_i(\vec{x}_i=0,\vec{y}_i)\equiv 0  \qquad
\phi_i(\vec{x}_i,\vec{y}_i=0)\equiv 0 \label{BC1}
\end{eqnarray}
\item[--] for a 3-body bound state they decrease exponentially in all the directions. In practice one can force them to vanish at sufficiently large distances
\begin{equation}\label{BC2}
\phi_i(x_i\geq x_{max},y_i\geq y_{max})=0
\end{equation}
\item[--] for an open  i+(jk) elastic or rearrangement scattering, the i-th Faddeev component
splits into the product of the two-body bound state wave function $\varphi_i(x_i)$ of the particle pair (jk) and
the scattering wave of particle i with respect to the center of mass of this pair $\chi_i(\vec{y_i})$:
\begin{equation}\label{BC3}
\phi_i(\vec{x}_i,\vec{y}_i)=\varphi_i(x_i)\chi_i(\vec{y_i})
\end{equation}

\item[--] for the break-up reactions at large values of the hyperradius $\rho=\sqrt{x_i^2+y_i^2} $ one overimposes to (\ref{BC3})  the behavior
\begin{equation}\label{BC4}
 \phi_i(\vec{x}_i,\vec{y}_i) ={e^{ik\rho}\over \rho^{1/2} }
\end{equation}
\end{itemize}

\bigskip
As mentioned in the Introduction, the original Faddeev equations, above formulated,
are not suitable for the Coulomb scattering problems.
The reason is that the right hand sides of eq.~(\ref{FE}) do not decrease fast enough
to ensure the decoupling of Faddeev amplitudes in the asymptotic region
and to allow  unambiguous implementation of the boundary conditions.
In order to circumvent this problem, Merkuriev~\cite{Merkuriev_80,Merkuriev_81} proposed to split
the Coulomb potential $V$ into two parts, $V=V^s+V^l$, by means of some arbitrary cut-off function $\eta$:
\begin{equation}
V^s(x,y) = V(x)\eta(x,y) \qquad
V^l(x,y) = V(x)[1-\eta(x,y)]
\end{equation}
One is then left with a system of equivalent equations
\begin{equation}\label{MFE}
 \left[E-H_{0}-W_{i}-V^s_{i}\right]\Phi_{i}=
V^s_{i} \left[\Phi_{j}+\Phi_{k}\right]
\qquad   W_{i}=V^l_{i}+V^l_{j}+V^l_{k}
\end{equation}
with the right hand side containing only short range  contributions ($V_s$)
and with a 3-body potential ($W_{\alpha}$) in the left hand side to account for  the long range parts.
As it will be demonstrated in the following, this approach was found to be very efficient
in calculating the positron-positronium (Ps) and positron-hydrogen (H) cross sections~\cite{KCG_PRA_92,KCG_PRA_95}.

\bigskip
The main advantage of the Faddeev formulation is to allow a proper description of the different
asymptotic states, each of them well separated by the corresponding Faddeev component and
easily described  in terms of their natural Jacobi coordinates. It is clear
 that the implementation of multichannel boundary conditions
listed  in (\ref{BC1}--\ref{BC4}) is impossible by means of one single function.
Let us just however mention that this is not the only reason for using this approach, after all
when working in momentum space there is no any need to implement any kind of boundary conditions.
In fact there are also deeper reasons related to the mathematical structure of the underlying
Lipmann-Schwinger like integral equations. The reader interested can take benefit in consulting~\cite{Fad_67,MF_Book}.

Equations~(\ref{MFE}) allow to treat  any physical non-relativistic quantum three-body problem
with the only exception of break-up reaction with at least  one attractive Coulomb interaction, despite numerous attempts~\cite{ES_PRA67_03}.

Last but least, this formulation is also advantageous in the numerical calculations. Indeed, the Faddeev
decomposition takes benefit of  the symmetry properties of the system.
As a consequence, t he Faddeev components have simpler structure than the total wave function itself and are
therefore easier to interpolate numerically.

\bigskip
The above described formalism has been extended to account for particles with spin,
interacting via three-body and/or non local forces. Few years after their formulation, the
Faddeev equations were generalized to an arbitrary number (N) of particles by Yakubovsky~\cite{Yaku_67}.
However their numerical solution is still limited to N=4 case and
for the energies below the first three-body break-up threshold. In this manuscript
we will not present the formalism of Yakubovsky equations and their numerical
implementation techniques, interested reader may refer to~\cite{Thesis}. Still in
sections~\ref{Cold}-\ref{He_mers} we present some results, which were obtained by
solving four-particle Yakubovsky equations.

One should also mention that, although we have described Faddeev approach in coordinate space
in terms of coupled differential equations, the equivalent formulation is
available in momentum space using coupled integral equations. Both formulations have its
own advantages, namely higher energy scattering is easier to treat in momentum space~\cite{GWH_PR_96,Gloeckle_book},
while configuration space is more convenient for Coulomb related problems~\cite{MF_Book,KCG_PRA_92,LC_FBS31_2002,Papp_PRL94_05}.

\section{Charged particle interaction with H}\label{p_H}

An interesting atomic physics problem is related to the  polarization of neutral atoms.
In the vicinity of a charged particle $X^{\pm}$,  the electronic density of a neutral atom $A$  polarizes, giving rise to a central long range
$X^{\pm}-A$ interaction which behaves asymptotically like
\begin{equation}\label{V_P}
V_P(r)=-{1\over2}{ \alpha_d\over r^4}
\end{equation}
where $\alpha_d$ is the dipole polarizability, characteristic of the atom ($\alpha_d=9/2$ for H in the $m_p\to\infty$  limit).
This interaction is very shallow but if the charged particle is heavy enough to reduce the repulsive kinetic energy, it generates
a rich spectrum of bound and resonant states.
This can be qualitatively illustrated  in a simple two-body approach by assuming an $X^{\pm}-A$  two-body interaction of the form (\ref{V_P}) with the replacement~\cite{MM_87}:
\[   \alpha_d \to  \alpha (r)= \alpha_d- \frac{2}{3} e^{-2r} \left( r^5 + \frac{9}{2}r^4 + 9r^3 +\frac{27}{2}r^2 + \frac{27}{2}r    + \frac{27}{4}\right) \]
This particular choice of $\alpha(r)$, valid for scattering on $H$ atom,  regularizes (linearly) the polarization potential (\ref{V_P}) at short distances, keeping also  the right asymptotic behavior.

\begin{figure}[h]
\begin{center}
\begin{minipage}{7.5cm}
\begin{center}\mbox{\epsfxsize=7.cm\epsffile{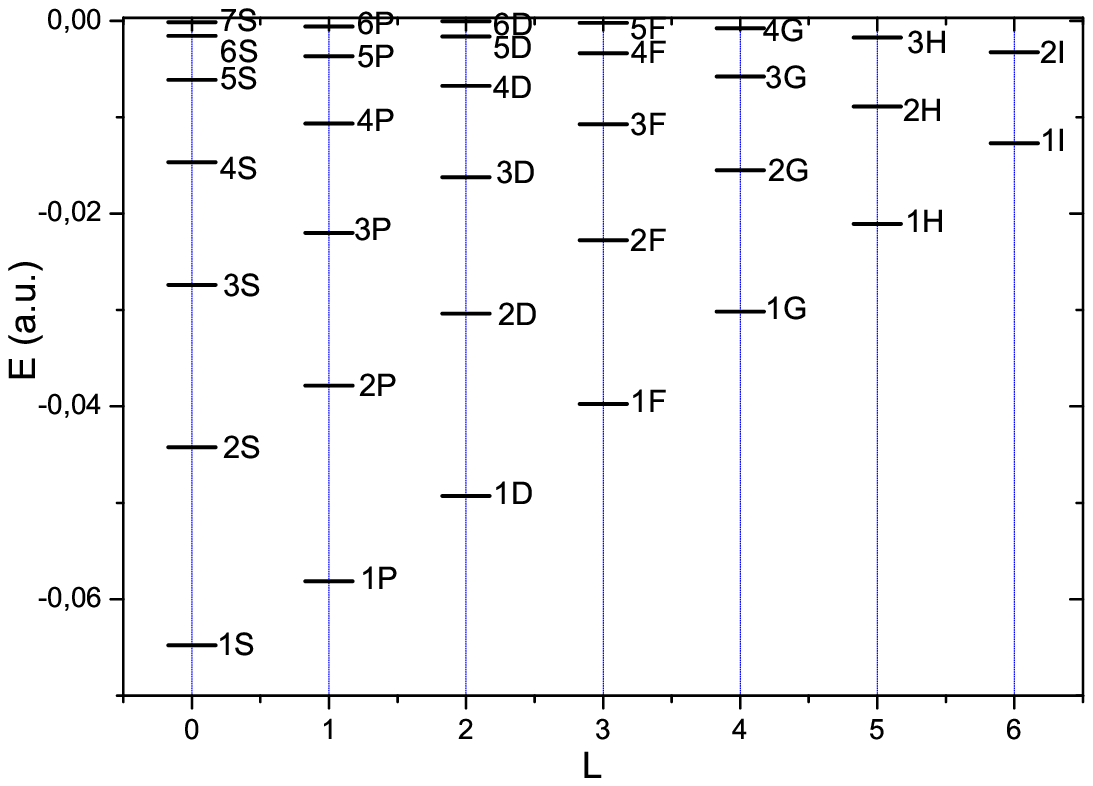}}\end{center}
\caption{Bound state spectrum (atomic units) for $\pi$-H in a two-body approach with a regularized polarization potential (\ref{V_P}).}\label{B_piH_2c}
\end{minipage}
\hspace{0.5cm}
\begin{minipage}{7.5cm}
\begin{center}\mbox{\epsfxsize=7.cm\epsffile{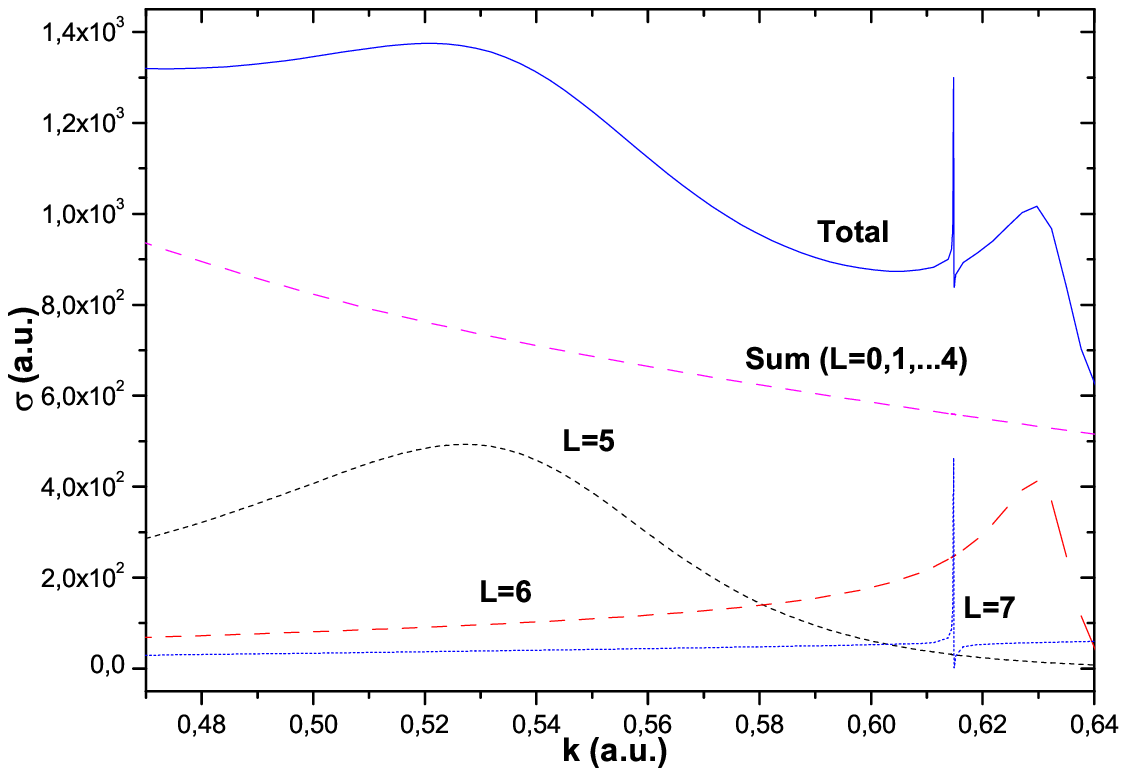}}\end{center}
\caption{Resonant $\mu-H$ states in the same two-body approximation as in  figure  \ref{B_piH_2c}.}\label{Res_muH_2c}
\end{minipage}
\end{center}
\end{figure}

We have plotted in Figure~\ref{B_piH_2c} the bound state spectrum of the $\pi-H$ system thus obtained.
It contains a large number of states up to $L=6$. In the case of a positively charged particle, $X^+$, these states
would lay  below the rearrangement threshold $p+(e^-X^+)$.
Figure~\ref{Res_muH_2c}  represents the $\mu-H$ elastic cross section
obtained in the same  approximation, displaying very narrow resonances in  high angular momentum states~\cite{LC_FBS31_2002}.

In view of having reliable quantitative predictions, an exact calculation must be however performed, taking into account the internal structure
of the atom.
In the present Few-Body {\it state of the art},  this is only possible for the H and He atoms.
The simplest case for which the calculations
can be performed exactly is the scattering on atomic hydrogen. This constitutes a genuine three-body problem
and a challenge for the Few-Body community.

\begin{figure}[h]
\vspace{-1.5cm}
\begin{center}
\epsfxsize=10.cm\epsfysize=8.cm {\epsffile{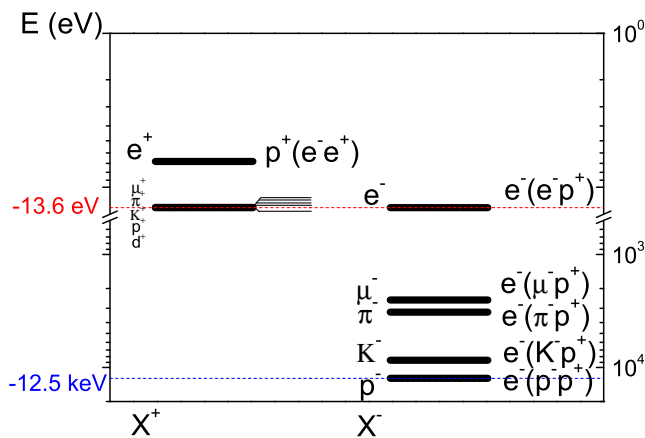}}
\caption{Different three-body reactions with atomic hydrogen and the existing particles.}\label{Reactions}
\end{center}
\end{figure}

\begin{figure}[h]
\begin{center}
\begin{minipage}{7.5cm}
\begin{center}\mbox{\epsfxsize=7.cm\epsffile{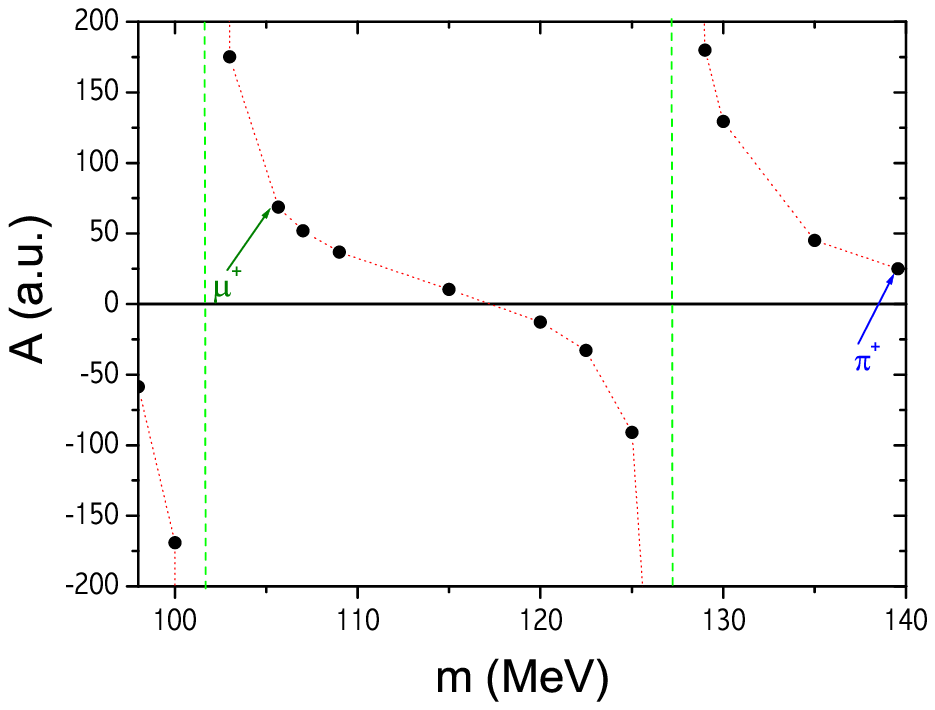}}\end{center}
\caption{$X^+-H$ scattering length (in atomic units) as a function of the projectile mass (in MeV).
The values corresponding to physical particles are indicated by arrows.}\label{A_m}
\end{minipage}
\hspace{0.9cm}
\begin{minipage}{7.5cm}
\begin{center}\mbox{\epsfxsize=7.cm\epsffile{sig_pH_A_CRAS.eps}}\end{center}
\caption{Cross section for the L=0 pH scattering in pp spin triplet state as a function of the energy (in atomic units).
The body results (filled black square) are compared to those (red solid line) given by the two-body Landau potential~\cite{LL_3} modified in order to reproduce the binding energy the first excited state.}\label{Sig_pH}
\end{minipage}
\end{center}
\end{figure}

We have first considered the case of positively charged particle $X^+$ with $m_{X^+}\leq m_p$, for they have at zero energy  a single open channel
corresponding to elastic scattering (see Figure \ref{Reactions}).
The first exact three-body solutions of the $X^+(e^-p^+)$  have been obtained in~\cite{LC_FBS31_2002}. We display in Figure \ref{A_m}
the $X^+-H$ scattering length as a function of the mass of the projectile $m_X$.
The values corresponding to physical particles  ($\mu,\pi$) are indicated by arrows.
As one can see from this figure, the scattering length
is divergent for some values  of the incoming particle mass, indicating the appearance of an additional bound state in the ($X,p,e$) system.

\bigskip
Of special interest is the scattering of protons. Not only for they constitute experimentally  the most
 accessible physical particle but also because it corresponds, by chance, to one of the resonant cases displayed in Figure \ref{A_m} for smaller projectile masses.

Indeed we have computed the $p-H$ scattering length for the S-wave symmetric (pp spin singlet) and antisymmetric (pp spin triplet) states.
For the symmetric case we found $a_s=-29.3$ a.u. while the antisymmetric one provided the value of $a_t=750\pm5$ a.u.
The analysis of the nodal structure of the corresponding Faddeev amplitude
indicated that such a big value is due to the existence of a first excited bound state with extremely small binding energy.
By using the effective range expansion, its binding energy was found to be B=$(1.135\pm0.035) \times 10^{-9}$ a.u. that is. $\approx30$ neV.

This state can be also viewed as the first excited vibrational level $v=1$ of  $2p\sigma_u$  symmetry in the H$_2^+$ molecular ion.
A direct computation of this state using ad-hoc variational techniques~\cite{HBGL_EPJD_00}   confirmed its existence
and provided a more accurate value of the binding energy $B=1.085045 \times 10^{-9}$~\cite{EPL64_2003}.
Further work showed that it is stable with respect to the relativistic and leading order QED corrections. Taking
them into account, its binding energy  is only slightly modified and becomes $B=1.082247 \times 10^{-9}$~\cite{JPB37_2004}.

It is worth mentioning that this $H_2^+$  first excited antisymmetric state exists also in the so called Landau $p-H$ potential~\cite{LL_3}.
The latter consists in adding to the polarization term~(\ref{V_P}) a repulsive one due to the Pauli principle
between protons in the spin triplet state and reads (in atomic units)
\begin{equation}\label{VL}
 V_{L}(x) = \eta  \frac{2x}{e^{x+1}} - \frac{\alpha_d}{2 x^4}
\end{equation}
The total potential $V_L$ is regularized to a constant below  $x_c=2.5$. In its original formulation ($\eta=1$) it entails an excited  S-wave state, although with binding energy
two order of magnitude smaller than the exact three-body value  and a $p-H$ scattering length  consequently larger.

To our knowledge the $H_2^+$ first vibrational  $2p\sigma_u$ state above described constitutes the most weakly bound natural molecule ever predicted \footnote{It is however possible to prepare
arbitrarily weakly bound systems suitably adjusting external magnetic fields, like for instance in  \cite{Hulet_PRL_2009}}.
A direct computation provides a root mean squared radius $R=270$ a.u. and its wavefunction  has still sizeable values  well beyond
1000 a.u.. That makes the state extremely unstable against any kind of perturbation.
The low energy $p-H$ scattering will be however totally dominated by the existence of this nearthreshold state.
Corresponding S-wave   elastic $p-H$ cross section is shown in Figure \ref{Sig_pH}.
The Faddeev results (black filled squares) are compared to those (red solid line) obtained with the Landau potential (\ref{VL}),
whose parameter was adjusted to $\eta=0.9254$ in order to reproduce the exact 3-body binding energy.
The huge values of the cross section are a consequence of the large triplet scattering length $a_t$. Notice that the resonant
region is however limited to $E<10^{-7}$  a.u.. Below this energy range, a proton approaching an hydrogen atom
will feel an object of nanoscopic size.

\begin{figure}[h]
\begin{center}
\begin{minipage}{8.cm}
\begin{center}\mbox{\epsfxsize=8.5cm\epsffile{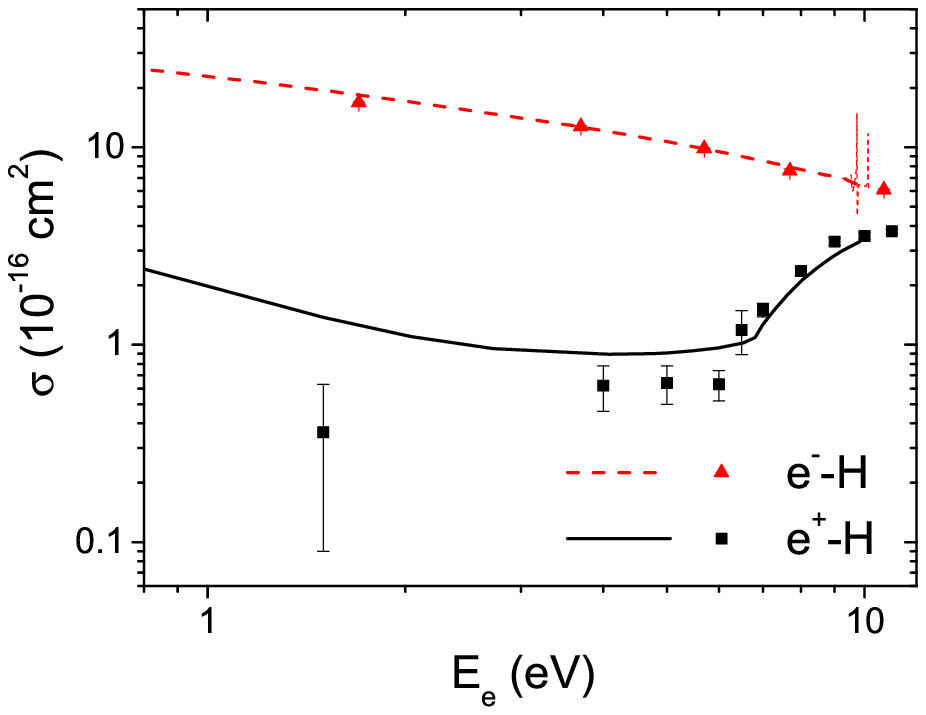}}\end{center}
\caption{$e^\pm$-H scattering cross section below H($n=2$) threshold: calculated
values are compared with experimental ones~\cite{Zhou}.}\label{epm_H}
\end{minipage}
\hspace{0.9cm}
\begin{minipage}{8.cm}
\begin{center}\mbox{\epsfxsize=8.5cm\epsffile{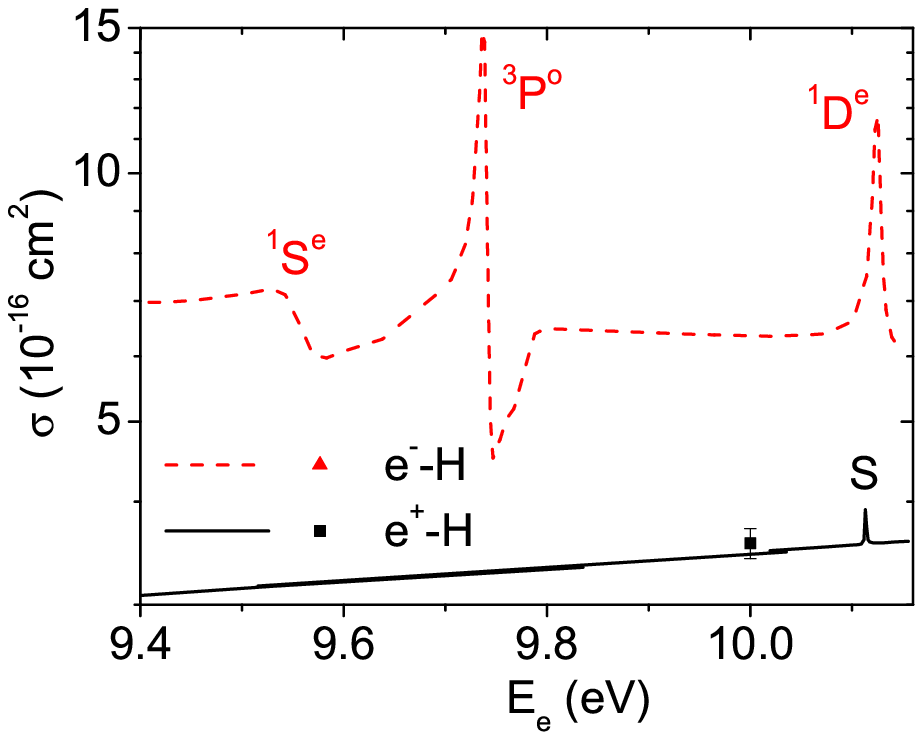}}\end{center}
\caption{A zoom in the resonance region of $e^\pm$-H scattering close to H($n=2$) threshold.}\label{Zoom_epm_H}
\end{minipage}
\end{center}
\end{figure}

\bigskip
Electron and positron scattering on Hydrogen atoms in their ground state
presents one of the rare three charge structures for which experimental data
are available. In fig.~\ref{epm_H} we compare our calculated
total scattering cross sections with the measured ones~\cite{Zhou}, for  $e^\pm$ energies below H($n=2$) threshold ($E_2\approx 10.2$ eV).
One may see an excellent agreement between measured and calculated values.
The only exception is in the $e^+$-H scattering at very low energies  (E$\approx$ 1eV), where
the experimental  result has a big error bar, probably due to the strong positron annihilation probability.

We have displayed in figure \ref{Zoom_epm_H} a zoom of the resonance region, close
to the H($n=2$) threshold. One may identify three well-known narrow
resonances in  $e^-$-H system, which are also confirmed
experimentally \cite{Resonances_eH_72,Resonances_eH_76,Resonances_eH_86}. One S-wave resonance is visible in  $e^+$-H system.

\bigskip
If the incoming particle is heavier than the proton, e.g $X^+=d$, the rearrangement process
\begin{eqnarray*}
 X+ (e^-p)  &\to&  p+(e^-X^+)
\end{eqnarray*}
is kinematically open and even at zero kinetic energy  we  are left with a two  channel problem.
By increasing the scattering energy, the number of open 1+2 channels before the 3-body breakup threshold becomes infinite.

The situation is even more interesting in the case of an incoming negatively
charged $X^-$ which presents even at zero incoming energy a large numbers of  open channel corresponding to Coulomb ($X^-,p$) ground and excited states (see Figure \ref{Reactions}).
All of them provides a very rich variety of phenomena, most of them
remaining unexplored both from the theoretical as well as from the experimental point of view.

\section{Cold molecules \label{Cold}}

An important simplification of the Faddeev equations may take place if one considers a
system of identical particles. In this case, the three Faddeev equations,
as well as the corresponding components, become identical and therefore the full
solution of the problem is obtained by solving only one of them.
Such symmetrized equation may be applied to study a system of
identical bosons. This is the simplest few-body problem, nevertheless revealing genuine peculiarities of quantum many-body systems.

Two of the striking phenomena manifested in few body physics are the Thomas collapse~\cite{Thomas} and
the so called Efimov states~\cite{Efimov_70,Efimov_79}.

The first of these two effects, the Thomas collapse, attests that  an $N>2$ boson system will shrink into a state
with an infinite binding energy, if the pairwise particle interaction is attractive and of zero-range.

The second phenomenon, i.e. Efimov states, manifests in a series of weakly bound three-body states,
which move below the two-body threshold by slightly increasing the two-body attractive interaction.
These states can appear in a realistic physical system with
finite range interaction, if one of its two-body subsystems is S-wave resonant, i.e.
when the two body scattering length ($a_0$) turns to be much larger than the interaction range
($r_0$). The interest in Efimov states has been fueled lately by the important
progress of experimental low temperature physics and particularly by the recent discovery of an Efimov state in caesium atoms~\cite{Nature_06}
and the subsequent, even more conclusive, experimental works of refs. \cite{Ef_1,Ef_2,Ef_3,Ef_4}.

Efimov physics involves systems having largely separated length (or momenta) scales
and therefore turns to be a natural laboratory to test Effective field theories (EFT).
To analyze Efimov states it is then natural to introduce EFT expansions
in terms of two small dimensionless parameters $r_{0}/a_{0}$ and $kr_{0}$, $k$ being a characteristic center-of-mass (c.m.) momenta.

It is a common believe that, at the leading order of EFT, all the low-energy
properties of a three-boson system are set by two parameters: one two-body
parameter -- like two-boson (dimer)  binding energy or scattering length -- and one
three-body parameter --  like three-boson (trimer) binding energy or particle-dimer scattering
length~\cite{MC_HAMMER,Frederico_02}.
This means that whenever in a three-body system the particle interdistance $R$ satisfies the Efimov condition,
$r_{0} \ll R \ll |a_{0}| $, its low-energy properties
must be interaction independent and be governed by the only two low energy  aforementioned parameters.

More recently, it has even been demonstrated that the properties of
the four-boson system (tetramer) are determined by the same two parameters and no
additional four-body scale is required to establish universal relations
between three- and four-boson observables~\cite{Platter,stecher}. These  universality
studies have been also extended to fermionic three- and four-body systems,
recovering series of phenomenologically
observed  correlation rules in multi-particle system~\cite{Platter_N}.
Most of these relations have been derived relying on  purely attractive
 (and often contact) two-particle interactions, while $N>2$
 systems are balanced by introducing repulsive three-particle force to prevent Thomas collapse .

We present  in this section  a slightly different quantum-mechanical approach, based on a family of short range
separable, rank-1 and rank-2, interactions having the same two-boson scattering length.

These potentials have been derived somehow mimicking interaction between $^{4}$He atoms, i.e.,
choosing the boson mass  $\hbar^{2}/m=$ 12.12 K$\cdot${\AA}$^{2}$
and fixing the two-boson scattering length to $a_{0}=$104.0 {\AA}.
Nevertheless one can not pretend with these results a realistic description of
$^{4}$He multimers, since  these separable potentials do not retain
well known properties of the effective interaction of two inert neutral atoms, like
strong repulsion at the origin and weak van-der Walls attraction in the asymptote.

The family  of two-boson  separable (and thereby nonlocal) potentials we use have the form:
\begin{equation}
v = \sum_{ij}^{n_r} |g_i\rangle \lambda_{ij} \langle g_j|
\end{equation}
where $n_r$ is the rank of the considered potential.
For simplicity, these potentials are restricted to $S$-wave only. In close analogy with most of the
EFT potentials~\cite{Platter}, the form factors $|g_i\rangle$  are chosen as Gaussian.
In the configuration-space they are given by
\begin{equation}
\langle r |g_i\rangle = (\Lambda_i/\sqrt{2})^3 \, e^{-(\Lambda_i r/2)^2}
\end{equation}

The rank-1 potential is defined by two parameters: $\lambda_{11}$ and $\Lambda_1$.
To preserve the two-body scattering length at $a_{0}=$104.0 {\AA}, these two parameters must
satisfy $\lambda_{11} = [\pi m (1/a_0 - \Lambda_1/\sqrt{2\pi})/2]^{-1}$.

The rank-2 potential is obtained by choosing $\Lambda_j = j\Lambda$
with $\Lambda$  in the range $[0.133,0.305]$ \AA$^{-1}$, i.e.  $13.8 < a_0 \Lambda < 31.7$,
and determining  the remaining three strength parameters $\lambda_{ij}$ (we assume $\lambda_{12} = \lambda_{21}$)
by a fit of the calculated observables (with a typical  four digit accuracy):  dimer binding energy $B_2=$1.318 mK, scattering length  $a_{0}=$104.0 {\AA}
and two-particle scattering phase-shifts up to about $50 \, B_2$ c.m. energy.
The broad scattering energy interval  included in the fit,
roughly corresponding to the natural energy scale  $\hbar^{2}/mr_0^2 $~\cite{Platter},
guarantees that in the effective range expansion not only $k^2$ but also higher order terms are  well retained.
Note that the phaseshift set we have used to fix the parameters of the rank-2 potential
differs from the one obtained by realistic interaction (for instance  our effective range   $r_0 = 15.0$ \AA is about two times larger).
\begin{figure}[tbp]
\epsfxsize=10.0cm
\epsffile{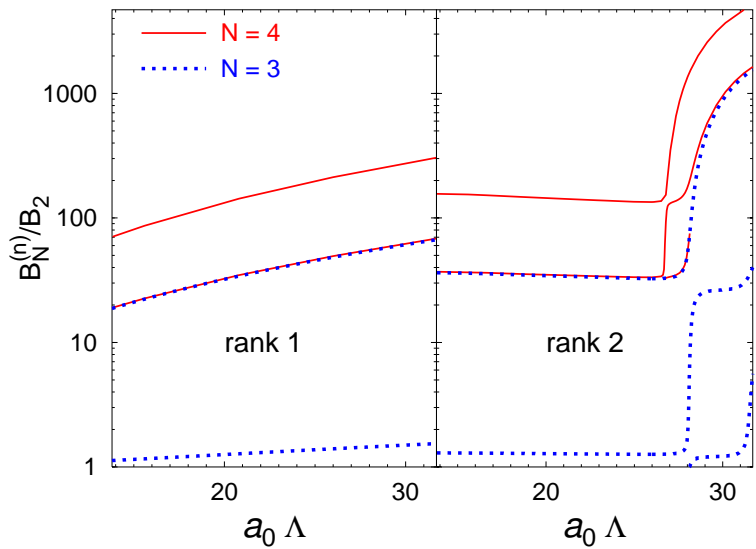}
\caption{\label{fig:b-l}
Evolution of the three- and four-boson binding energies with the Gaussian cutoff $\Lambda$.}
\end{figure}

The dependence of the binding energies of L=0$^+$ three- and four-boson states
as a function of the Gaussian cut-off $\Lambda$~\footnote{One should notice
that the described potentials do not support L$\neq$0$^+$ rotational states, neither for trimers
nor for tetramers} is displayed in fig.~\ref{fig:b-l} for both rank-1 and rank-2 potentials.
By binding energy $B_n$ of a n-particle multimer we denote the difference between the sum of the
n free constituent particles masses and the total multimer mass.

\begin{figure}[tbp]
\epsfxsize=10.0cm
\epsffile{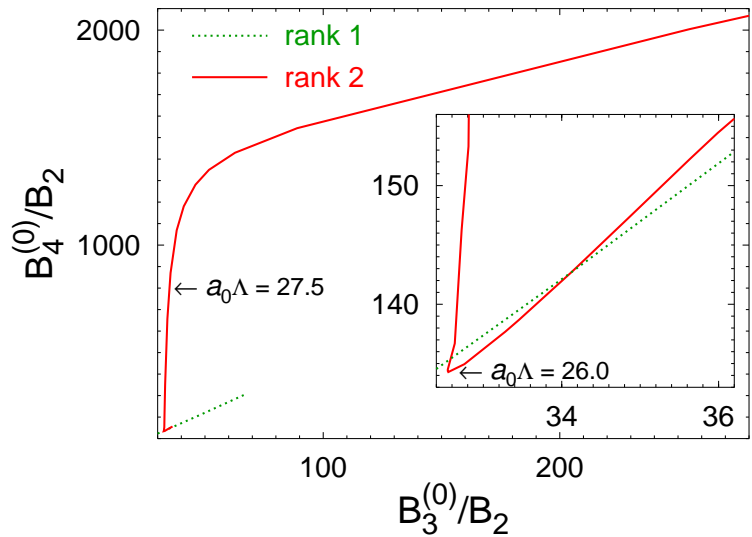}
\caption{\label{fig:b4}
(Color online)
Correlation between ground state binding energies of trimer and tetramer.}
\end{figure}

The rank-1 results are consistent with the  EFT predictions, based on contact 2- and 3-body
interactions. There exist two trimers, the ground state which is strongly
bound (its binding energy exceeds the dimer one by  more than a factor 10), and the
first excitation which remains close to the dimer threshold.
There are also two tetramers,
the ground state being strongly bound relative to trimer one and
the excited tetramer which consist in  one boson  weakly attached to the trimer ground state.
The binding energies of these trimer and tetramers
reveal almost a linear correlation pattern for rank-1  potential, see Fig.~\ref{fig:b4}, and are in agreement  with  the EFT results of ref. \cite{Platter_N}.

The situation is much more complicated when using the rank-2  potential.
The binding energies $B_N^{(n)}/B_2$ are slowly decreasing functions of the cut-off  $\Lambda$
in the range $a_0\Lambda < 26.0$.
In this region, all potentials support one ground and one
excited state for the trimer as well as for the tetramer, in agreement with the EFT approach.
Moreover the trimer and tetramer binding energies follow closely the correlation curves of rank-1 values.
However, for larger $\Lambda$ values, the rank-2 potential results show a nontrivial behavior despite the fact
that all the potential parameters $\lambda_{ij}$ depend smoothly on $\Lambda$.

First, at $a_0\Lambda = 26.5$, the binding energy of the excited tetramer $B_4^{(1)}$
detaches from $B_3^{(0)}$. It increases drastically and stabilizes for a moment
when it almost attains
the tetramer ground state binding energy $B_4^{(0)}$ that, with a tiny delay, also starts to increase rapidly.
At the same time, close to $a_0\Lambda = 26.75$, the second excited tetramer state appears with a
binding energy $B_4^{(2)}$ only slightly larger than $B_3^{(0)}$.
In this region of harsh tetramer variations, the trimer properties remain practically unchanged.

The trimers exhibit  qualitatively similar behavior, although at larger $\Lambda$ values, around $a_0\Lambda = 27.5$.
 The binding energies of the trimer ground and first excited state,
 $B_3^{(0)}$ and $B_3^{(1)}$,  start to grow very rapidly; the former increases much like $B_4^{(0)}$ did
while the latter stabilizes for a while.
As  $B_3^{(0)}$  increases, the second excited tetramer state disappears below the ground
trimer threshold at $a_0\Lambda = 28.14$, whereas the first excited tetramer stalls for a
while letting the trimer threshold $B_3^{(0)}$ to approach again.
Furthermore, at $a_0\Lambda = 28.12$ the second excited trimer state appears.
Unlike the strongly bound trimers, this state behaves like an Efimov one, i.e., it
slides under the dimer threshold if the two-boson interaction is made stronger.
The second excited trimer loses that property above $a_0\Lambda = 30.7$
where its binding energy  $B_3^{(2)}$, much like  $B_3^{(1)}$  at $a_0\Lambda = 27.5$, starts a phase of a rapid increase.
However, the Efimov property is recovered by the third excited trimer,
appearing at $a_0\Lambda = 31.7$.  It is interesting to note
that the second excited tetramer  shows a similar behavior, i.e. it also
disappears if the two-boson interaction is made stronger.

While rank-2 potential results seems to be rather confusing,
the explanation of such a nontrivial behavior of a few-boson system is rather simple.
For this aim one should simply understand the $\Lambda$-evolution of the nonlocal interaction.
The shape of $n_r=1$ potential is very simple, and can be represented by a single attractive
bell-shape well in $(r,r')$ plane, the width of the bell being constrained by the size
of the cutoff $\Lambda$. For  small $\Lambda$ values, the $n_r=2$ potential has also single
and rather flat attractive plateau in the $(r,r')$ plane. However for larger  $\Lambda$
values, this plateau splits into a rather complicated saddle-like surface with
two asymmetric attractive regions separated by two symmetric repulsive regions.
A narrow attractive region is formed at the origin and it deepens when increasing $\Lambda$,
whereas the second,  distant, attractive region is much wider and flatter.
Due to the large kinetic energies required to squeeze the particles into the narrow region at the origin,
it is energetically preferable for the multi-boson system to stay in the distant attractive plateau.
At this moment, the $n_r=2$ potential presents qualitatively the same features as $n_r=1$, having a single
attractive region. However, for some values close to  $a_0\Lambda = 26.0$, the
 attractive region at the origin becomes deep enough to accommodate the four-boson system, and
some kind of a phase transition occurs.  The trimer has only three pair interactions
compared to six for the tetramer. Therefore the corresponding phase transition in the three-boson system
takes place at larger  $\Lambda$ values, when the attractive region near the origin is even more deepened.

One can probably question the reality of such "exotic'' nonlocal potentials with several attractive regions.
Nevertheless, one must  admit that some realistic nucleon-nucleon potentials~\cite{entem:03a,doleschall:04a} have even more
complex structure than the one exhibited in our model.
Effective interaction between two complex molecules can however reveal even
much more complicated pattern.

The presence of finite  range ($r_{0}>0$) interactions does not guarantee that the Thomas-like collapse will not occur in multi-boson system.
The interaction can have a short-range or an off-shell structure that may be ignored with the low-energy probe. Such structure
can have no or little effect for low-energy observables of an $N$ particles system, but may be exploited in the
$N+1$ particle one which, having more interacting pairs, can compensate larger kinetic energies and
thus regroup itself into a shorter range domain.
This fact complicates the possibility to make universal predictions about
multi-particle system, in particular about its deeper lying states, based
only on the low energy properties of its subsystems, unless one is sure that the
eventual size of the few-body system satisfies the Efimov condition $R >>r_0$.
Of course, this condition is never satisfied a priori, although one
may expect that it must be valid for higher order excited states~\cite{Arnas_PRL}.

\section{$^{4}$He multimers}\label{He_mers}

In this section we turn our attention to a realistic description of small structures of $^{4}$He atoms.
The two electrons of the He atom close
the 1s shell, determining its spherical symmetry  as well as  its inert chemical properties.
Two He atoms repel strongly when approaching to each other, but a very weak Van der Vaals attraction
is exhibited at large distance, resulting in a shallow attractive pocket with a maximal depth of $\approx$10.9 K, centered at $R_{He-He}\thickapprox 3$ \AA.

This weak attraction is responsible for the fact that at very low temperatures both bosonic $^{4}$He but also fermionic $^{3}$He become liquid.
It is also responsible for the existence  of a loosely   bound $^4$He dimer, with binding energies of $B\approx1$ mK and sizeable
 $^4$He- $^4$He scattering length $a_0\approx100$ \AA, which has been experimentally observed few years ago~\cite{Spuch,Toennies}.

When trying to study theoretically N$>$2 $^4$He clusters, one is faced to a serious theoretical problem.
The standard boundary conditions in the Dirichlet form (\ref{BC1}), which imposes to
the FY components to vanish at the origin, turns to be impractical when considering structures of He atoms.
The strong hard-core repulsion
describing the inner part of the He-He potential, which dominates the interaction  between two He atoms at relative distance  $R_{He-He}\approx$2 \AA,
introduce severe numerical complications. The relevant attractive matrix elements which are responsible for the binding fade away in front of these
huge repulsive hard-core terms, thus causing  numerical instabilities.
On the other hand, such strong repulsion close to the origin physically, simply
reflects the fact that the two He atoms cannot get close  to each other, i.e.
closer than some relative distance $r=c$ inside the core.

The wave function of three He atom system, for instance, should vanish in the region of three-particle space $\mathbb{R}^{6}$, which is the interior of the
three multidimensional surfaces $x_{i}=c,$ where $x_{i}$
is the distance between the particle $j$ and $k$ defined in (\ref{x_i}). Therefore one way to attain
numerical stability would be to neglect the solution in the strong repulsion region.
In practice, as it has been shown by  Motovilov and Merkuriev~\cite{Motov}, the
presence of infinitely repulsive interaction at $x_{i}< c$,
can be formulated in terms of boundary conditions for the Faddeev components, which are implemented by setting:
\begin{eqnarray}
\left[ E-H_{0}-V(x_i)\right] \Phi_i(\vec{x}_i,\vec{y}_i)=0 & & \text{for } x_i<c, \cr
\Phi_i(\vec{x}_i,\vec{y}_i)=0 & & \text{for } x_i=c.
\end{eqnarray}
A similar approach may be applied also to condition Faddeev-Yakubovsky equations for
four-body system, see~\cite{Motov}.

\bigskip
We compare  in Table~\ref{tab:he-mers} the calculated properties of three and four $^{4}$He atom
structures. The binding energies of the ground and excited states (in mK) together with
atom-dimer and atom-trimer scattering lengths (in \AA) are listed.
The realistic model is based on the interaction between two He atoms developed
by Aziz and Slaman~\cite{Aziz}, popularly referred to as LM2M2 potential.
These realistic calculations are compared to the values obtained using
the rank-1 potential from the previous section, whose parameters ($\lambda,\Lambda$) have been adjusted in order to reproduce
the  dimer and ground state trimer binding energies  obtained with the LM2M2 potential.
One may see that the rank-1 approximation reproduces well the excited trimer binding energy $B_3^{(1)}$
and the atom-dimer scattering length $a^{\{2-1\}}_0$. However the description of the four atom system
deteriorates considerably. Probably the main reason is the difference of
the off-energy shell properties between these two interaction models: a purely attractive rank-1 potential and
the LM2M2 one containing a strong hard core region.
The four boson system, being a more compact structure,
test more strongly the hard core region than the three body one, which is also reflected by the
fact that the rank-1 potential provides more binding than the LM2M2 one.

\begin{table}[tbp]
\caption{\label{tab:he-mers}
\protect Predictions for the $^{4}$He multimer properties.
Binding energies (in mK) and scattering lengths  (in \AA) obtained using realistic LM2M2 interaction and rank-1 separable model of section \ref{Cold}.}
\begin{ruledtabular}
\begin{tabular}{|l|cccccc|}
Pot.   & $B_3^{(0)}$ & $B_3^{(1)}$ & $B_4^{(0)}$ & $B_4^{(1)}$-$B_3^{(0)}$ &$a^{\{2-1\}}_0$ &$a^{\{3-1\}}_0$  \\\hline
LM2M2  & 126.39  &  2.2680 & 557.7   & 1.087   & 115.56 & 103.7 \\
rank-1 & 126.39  &  2.2758 & 597.9   & 3.16    & 114.77 & 67.38 \\
\end{tabular}
\end{ruledtabular}
\end{table}

It is worth mentioning that a direct calculation of the  $^{4}$He tetramer excited state still represents a challenging numerical task.
This state is very weakly bound and its wave function  extend over several hundreds of \AA.
One is therefore forced to use a very large and dense grid, ensuring at the same time enough accuracy
to trace the small binding energy difference with respect to the trimer ground state.
Nevertheless the vicinity of the tetramers excited state to the  $^{4}$He- $^{4}$He$_3$
continuum makes possible the extraction of its binding energy from the low-energy scattering results as explained in~\cite{LC_PRA73_2006}.

\section{Conclusion}\label{Conclusions}

The Faddeev-Yakubovsky equations provide   a rigorous quantum mechanical formulation
of the non relativistic few particle problem,  taking into account all the possible asymptotic states of the system.
First formulated in momentum space and for short range pair-wise  interactions, they
have been  generalized  to accommodate three-body  forces as well as long range Coulomb potentials.

Aside from its well defined mathematical structure, the Faddeev-Yakubovsky formalism turned to be very useful in numerical calculations.
A large variety of problems in hadronic, nuclear, atomic and molecular physics have been solved since its appearance.
Still limited to N=3 and 4 systems, they can be in principle extended to an arbitrary number of particles.
They provide very accurate results for the bound state problems
but the genuine feature of this formalism lies on its ability to treat scattering and
bound states on the same footing, thus enabling a complete and consistent analysis of the few-body system under consideration.

The configuration space formulation of Faddeev-Yakubovsky equations, made possible once the boundary conditions are well established, is the best adapted
to deal with atomic physics problems.
Some selected examples of them have been presented in this contribution covering aspects of cold atomic and molecular physics.
We have considered  the scattering of positively charged particle on atomic hydrogen, systems of cold atomic
molecules and bound and scattering states of $^4$He atoms.
In particular, we would like to emphasize the result concerning  the $p-H$ scattering:
the huge scattering length in the proton-proton spin triplet state $a_t\approx750$ atomic
units.

Even by restricting ourselves to the 3- and 4-body cases, a large variety of unsolved problems remains to be considered.
Among them we would like to mention:
\begin{itemize}
\item  The Coulomb reactions including break-up and/or many open channels, like those presented in Figure \ref{Reactions}
\item  The many  interesting antiproton physics processes, like $\bar{p}+d$, $\bar{p}H\to e^- +(\bar{p}p)^*$, $\bar{p} + (e^+e^-) \to \bar{H}+e^- $
\item  Four-body Coulomb problems, like first principle calculations of positron-positron scattering
\item For-Body break-up reactions involving 1+2+1 and 1+1+1+1 particle channels
\end{itemize}
They constitute altogether a very rich and challenging program   for the coming years, both from the theoretical and experimental physics  point of view.

\begin{acknowledgments}
This work was granted access to the HPC resources of IDRIS under the allocation 2009-i2009056006
made by GENCI (Grand Equipement National de Calcul Intensif). We thank the staff members of the IDRIS for their constant help.
\end{acknowledgments}

\end{document}